\documentclass[final,1p,times]{elsarticle} 
\usepackage{graphicx}
\usepackage{amssymb} 
\usepackage{amsthm} 
\usepackage{lineno}

\usepackage{url}

\newcommand{\Rcp}{$R_\mathrm{CP}$}
\newcommand{\pT}{$p_\mathrm{T}$}
\newcommand{\sNN}{$\sqrt{s_{{}_\mathrm{NN}}}$}

\journal{Nuclear Physics A} 

\begin{document}

\begin{frontmatter} 

% Your Title - please insert
\title{Spectra and nuclear modification factor of charged hadrons produced in lead-lead collisions at \sNN${}=2.76$\,TeV with the ATLAS detector at the LHC}

%% Single author (and collaboration) - please insert
\author{Petr Balek (for the ATLAS Collaboration)}
%\fntext[col1] {A list of members of the EMPIRE Collaboration and acknowledgements can be found at the end of this issue.}
\address{Charles University in Prague, Prague, Czech Republic}

%% Multiple authors
%\author[auth2]{Marcus Junius Brutus}
%\address[auth1]{Somewhere, Rome}
%\address[auth2]{Somewhere else, Rome}

\begin{abstract} 
The measurement of charged particle spectra in heavy ion collisions is a way to study properties of hot and dense matter created in these interactions. The centrality dependence of the spectral shape is an important tool to understand the energy loss mechanism. The ATLAS detector at the LHC accumulated integrated luminosity equal to 0.15\,nb${}^{-1}$ of lead-lead data at 2.76\,TeV per nucleon-nucleon pair. Due to the excellent capabilities of the ATLAS detector, and its stable operation in 2010 and 2011 heavy ion physics runs, these data allow measurements of the charged particle spectra and their ratios in different centrality bins over a wide range of transverse momenta  (0.5--150\,GeV) and pseudorapidity ($|\eta|<2.5$). The measured ratio central to peripheral events shows a suppresion by a factor of 5 at \pT${}=7$\,GeV. At higher \pT\ the ratio increases.
\end{abstract} 

\end{frontmatter} % do not change

%% linenumbers are useful for reviewing process
%\linenumbers

\section{Introduction}

The first results from the LHC experiments showed that the jets emerging from the hot dense matter created in heavy ion (HI) collisions do not retain their full energy \cite{aj_atlas}. Measurement of fully-reconstructed jets by the ATLAS experiment \cite{atlas} revealed a factor of two suppression of jet yields in central collisions with respect to peripheral ones \cite{jet_atlas}. Results from the LHC experiments (e.g. \cite{jet_paper}) do not show significant modification of the fragmentation function for large longitudinal jet momentum fractions carried by the charged particles. Measuring inclusive hadron production at high transverse momentum (\pT) originating from parton fragmentation is another way to study the mechanism by which hard partons traversing \mbox{the medium loose their energy.}

\section{Analysis}
The suppression in hadron production can be measured as a ratio of yields per nucleon-nucleon interactions in "head-on" HI collisions scaled by the number of nucleon-nucleon interactions ($N_\mathrm{coll}$) relative to those measured in p+p collisions. The p+p collisions can be approximated by the most glancing HI collisions scaled by $N_\mathrm{coll}$ as well \cite{confnote}. This ratio is \mbox{referred as \Rcp.}

This analysis uses data collected with the ATLAS detector in the 2010 and 2011 Pb+Pb runs with \sNN${}=2.76$\,TeV. After run and event selections, approximately 5.1$\cdot$10${}^{7}$ minimum bias (MB) events ($L_{int} \approx 7\mu$b${}^{-1}$) are analyzed in the 2010 data sample and a similar number of events in the 2011 MB data sample. In the 2011 data sample, hard probes (HP) events with $L_{int} \approx 0.14$nb${}^{-1}$ are analysed as well. HP events satisfy the extra requirement of an anti-$k_\mathrm{t}$ \cite{antikt} jet with the distance parameter of $R = 0.2$ and transverse energy ($E_\mathrm{T}$) of at least 20\,GeV estimated online.
%These events satisfy requirement for the presence of anti-$k_\mathrm{t}$ \cite{antikt} jet with the distance parameter of $R=0.2$ and with online estimated transverse energy ($E_\mathrm{T}$) of at least 20\,GeV.

The event centrality is estimated using the total $E_\mathrm{T}$ measured at the electromagnetic scale by the forward calorimeters (FCal) in the range $3.1<|\eta|<4.9$. The data are binned in 5\% and 10\% centrality bins. The mean number of nucleon-nucleon collisions $\langle N_\mathrm{coll} \rangle$ for each centrality bin, as well as its systematic uncertainties, is estimated using a Glauber Monte Carlo model \cite{glauber}. 

Monte Carlo simulations (MC) to study performance of the ATLAS detector are based on the HIJING \cite{hijing} and PYTHIA \cite{pythia} event generators. In the 2010 data analysis the environment of Pb+Pb collisions is simulated using $10^6$ HIJING events with embedded PYTHIA p+p hard scattering events. In 2011 data analysis the PYTHIA events are embedded into MB events recorded without noise suppression in the detector channel during the 2011 data taking.

Charged particle tracks are reconstructed using the ATLAS Inner Detector. Tracks are measured in the pseudorapidity interval $|\eta|<2.5$ and over the full azimuth using a combination of silicon pixel detector (Pixel), silicon microstrip detector (SCT), and a straw tube transition radiation tracker (TRT), all immersed in a 2\,T axial magnetic field. The minimum \pT\ of reconstructed tracks is 0.5\,GeV. Charged particles typically traverse three layers of silicon pixel detectors, four layers of double-sided microstrip sensors, and 36 straws. 

The track reconstruction settings used in Pb+Pb collisions were different from those used in p+p \cite{pptracks} and optimized for an increased occupancy in HI collisions. In the analysis a track is required to have at least 2 hits in the Pixel detector, one being in the innermost layer, and at least 7 hits in the SCT. Tracks are required to have zero Pixel and SCT holes. A hole appears on a track if a hit, predicted by the track trajectory, is not found in the detector layer.
The contribution from fake tracks (i.e. tracks composed of randomly associated hits in the detector layers) and secondary particles (decay products of e.g. B and D mesons) are reduced by requiring the significance of $d_0$ and $z_0 \sin \theta$, i.e. the distance of closest approach of the track to the event vertex in the transverse and longitudinal directions divided by the estimated uncertainty on this parameter for each measured track, to be less than 3. The $d_0$ and $z_0 \sin \theta$ parameters and their uncertainties are estimated by a vertex finding algorithm.

To reduce the amount of fakes at high \pT, tracks from the HP sample are required to be matched to anti-$k_\mathrm{t}$ jets with $R=0.2$. Furthermore, as the main interest of this analysis are spectra of charged hadrons, electrons and muons coming from electroweak decays of heavy bosons are subtracted from the measured spectra. This is done in all used samples.

\begin{figure}[ht]
	\centering
	\includegraphics[width=0.7\textwidth]{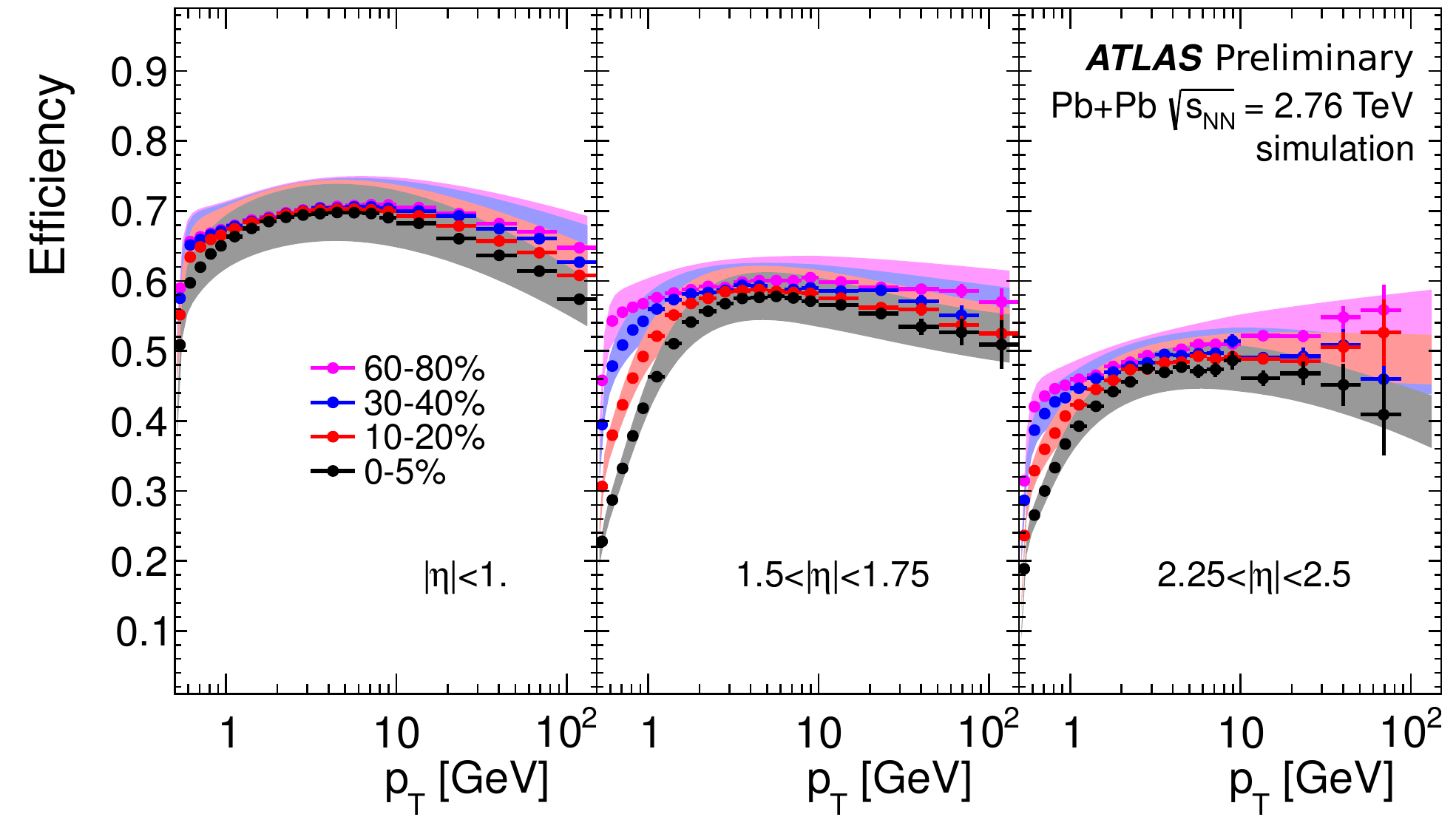}
	\caption{Tracking efficiency calculated for four centrality bins (0-5\%, 10-20\%, 30-40\% and 60-80\%) and three $|\eta|$ ranges ($|\eta|<1.0$, $1.5<|\eta|<1.75$ and $2.25<|\eta|<2.5$) derived from MC \cite{confnote}. The efficiency is a \pT\ spectrum of reconstructed tracks divided by a \pT\ spectrum of all truth tracks. Only particles with lifetime${}>0.3$$\cdot$$10^{-10}$\,s were used to produce efficiencies. The shaded bands show the fits of appropriate efficiencies with all their \mbox{systematic uncertainties combined.}}
	\label{effic}
\end{figure}

The data are corrected using MC for fake tracks and secondary particles, and for reconstruction inefficiency and momentum resolution. The fraction of fake tracks and secondary particles passing the tracking cuts among all tracks passing the same cuts is found to be around zero near $\eta=0$ for peripheral collisions and to increases both with increasing centrality and $|\eta|$. At the lowest measured \pT\ the fraction is up to 10\% (mainly contributed by secondaries) for central events and high $|\eta|$, and rapidly decreases with increasing \pT. At \pT${}\approx1$\,GeV the fraction is about 1\% and it is negligible above 4\,GeV. 
Corrections for the inefficiency of the track reconstruction and the momentum resolution are applied at once using bin-by-bin corrections. The efficiencies can be seen of Fig.~\ref{effic}. The track reconstruction efficiency is found to be highest in the region $|\eta|<1$ and decreases with increasing $|\eta|$. The efficiency decreases at low \pT\ as well as at very high \pT. In central collisions, the efficiency is lower by a few percent \mbox{compared to the peripheral collisions.}

The systematic uncertainties result from several contributions. Tracking cuts and vertex pointing cuts contribute up to 4\% each. The uncertainty on the efficiency correction contributes up to 20\% at high \pT\ and it is dominant in the spectra analysis. Uncertainties on momentum scale and material budget contribute up to 6\% and 5\% respectively. These uncertainties cancel out partially or completely when building the \Rcp. The dominant uncertainty of \Rcp\ results from the calculation of $\langle N_\mathrm{coll} \rangle$, which is up to 11.7\% for \Rcp\ \mbox{0-5\%/60-80\%}. Other uncertainties contribute less than 3\%, such as truth particle association or correction for fakes and secondaries.

\section{Results}

\begin{figure}[ht]
	\centering
	\includegraphics[width=0.7\textwidth]{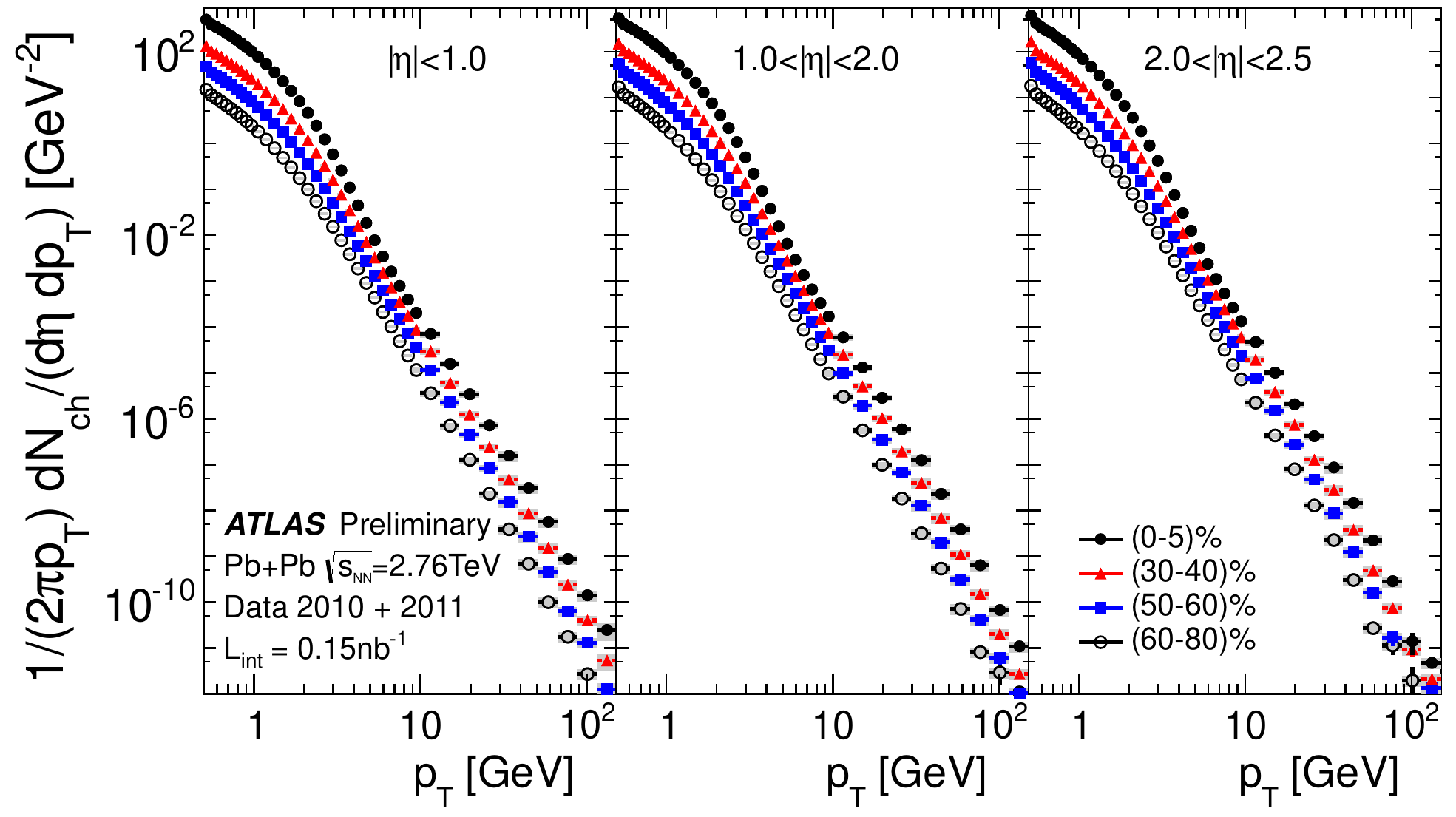}
	\caption{Fully corrected transverse momentum spectra for charged particles for four centrality bins and three $\eta$ ranges \cite{confnote}. The gray boxes show systematics uncertainties.}
	\label{spectra_exclusive}
\end{figure}

The final spectra are a combination of the MB spectra below \pT${}=30$\,GeV and HP spectra above. The spectra are normalized per event (per sampled event in case of HP spectra). The spectra as well as \Rcp\ in different $\eta$-regions are consistent with each other and can be combined. The corrected spectra are shown in Fig.~\ref{spectra_exclusive} for four centrality bins (\mbox{0-5\%}, 30-40\%, 50-60\% and 60-80\%), and three $\eta$ ranges ($|\eta|<1.0$, $1.0<|\eta|<2.0$ and $2.0<|\eta|<2.5$). 

\begin{figure}[ht]
	\centering
	\includegraphics[width=0.7\textwidth]{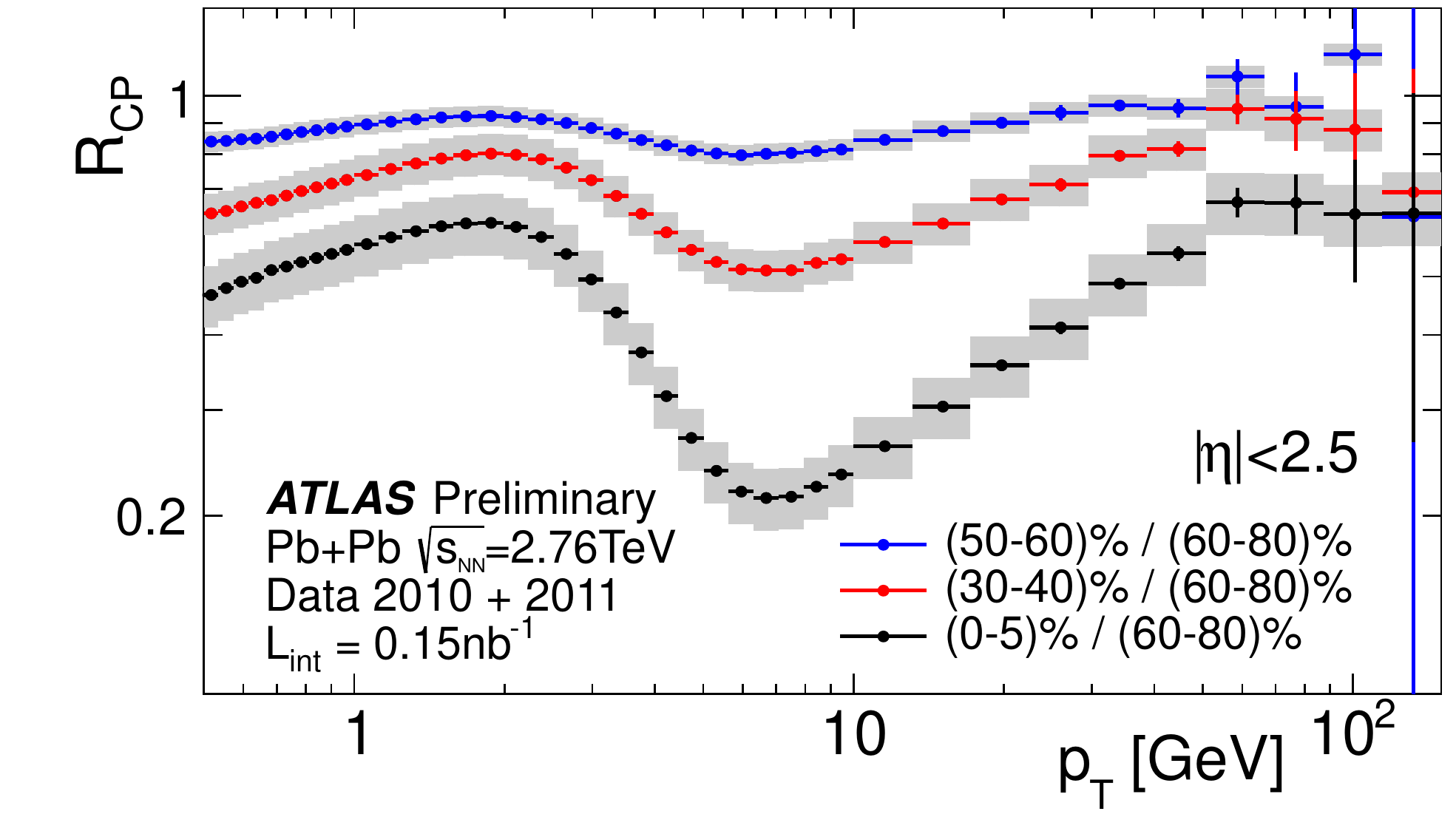}
	\caption{The \Rcp\ measured in $|\eta|<2.5$ in three centrality combinations \cite{confnote}. The statistical errors are shown with vertical lines and the overall systematic uncertainty at each point is shown with gray boxes.}
	\label{rcp}
\end{figure}

The \Rcp\ measured over the entire acceptance of the ATLAS Inner Detector tracking system as a function of track \pT\ is shown in Fig.~\ref{rcp}. It is shown for three centrality combinations: with 0-5\%, 30-40\% and \mbox{50-60\%} as the numerators and the common peripheral bin 60-80\% as the denominator. The measurement shows a strong suppression of the hadron production which increases with the centrality. The measured \Rcp\ for 0-5\% to 60-80\% centrality bins reaches a minimum at 7\,GeV and rises with \pT.

\section{Summary}

It was presented a measurement of inclusive charged hadron spectra in \sNN${} = 2.76$\,TeV Pb+Pb collisions and their ratios between central and peripheral bins. The measurement is performed in the range of $|\eta|<2.5$ and $0.5-150$\,GeV in different $\eta$ bins as the function of the \pT\ and the centrality of the collisions. Due to use of the jet trigger in the 2011 run, the statistics of high-\pT\ tracks is very good up to \pT${} \approx 70$\,GeV. The measurement of the \Rcp\ is limited by the statistics of the peripheral 60-80\% centrality bin. \mbox{Within this limited statistics the trend of the \Rcp} above \pT${} \approx 70$\,GeV is consistent with the size of the suppression measured in the jet analysis \cite{jetpaper}.

This research is supported by GAUK (projects nr.~435511, 500912 and 713612) and by \mbox{FP7-PEOPLE-IRG} (grant 710398).

%\section*{References}

\end{document}